\documentstyle[twocolumn,aps,prb]{revtex}
\input epsf
\newcommand{\cebsix}{CeB$_{6}$}
\newcommand{\celabsix}{Ce$_{x}$La$_{1-x}$B$_{6}$}
\newcommand{\TQ}{T$_{Q}$}
\newcommand{\TQH}{T$_{Q}$(H)}
\newcommand{\TN}{T$_{N}$}
\newcommand{\TK}{T$_{K}$}
\newcommand{\etal}{{\em et al.}}
\newcommand{\tm}{$^{TM}$}

\begin{document}
\twocolumn[\hsize\textwidth\columnwidth\hsize\csname@twocolumnfalse\endcsname 
\title{Magnetic Field Dependence of the Paramagnetic to the High 
Temperature Magnetically Ordered Phase Transition in \cebsix}
\author{Donavan Hall and Z. Fisk}
\address{National High Magnetic Field Laboratory\\
Tallahassee, FL 32306\\
}
\author{R. G. Goodrich}
\address{Department of Physics and Astronomy\\
Louisiana State University, Baton Rouge, LA 70803-4001\\
}
\date{\today }
\maketitle

\begin{abstract}
We have measured the magnetic field dependence of the paramagnetic to
high temperature magnetically ordered phase transition \TQH~ in \cebsix~ from 2
to 30 T using cantilever magnetometry.  It is found that the phase
separation temperature continuously increases in field with an
increasingly positive slope.  In addition, we find that measurements
in strong magnetic field gradients have no effect on the phase
transition.
\end{abstract}


\begin{center}
    Accepted for publication in \textit{Physical Review B} on Feb. 
    17, 2000
\end{center}

]



\section{Introduction}\label{sec:intro}
\begin{figure}[p]
	$$\epsffile{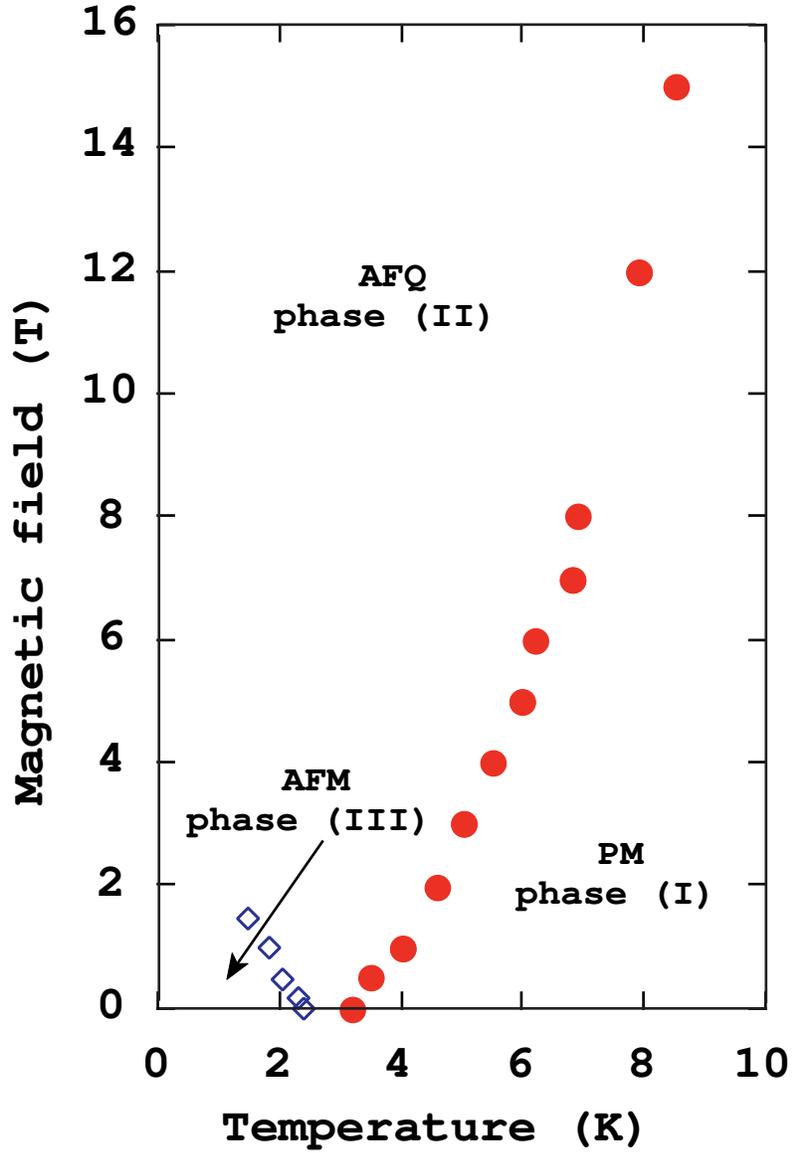}$$
\caption{The magnetic phase diagram of \cebsix~ exhibits three main
phases at zero field separated by two magnetic ordering temperatures:
the quadrupolar ordering temperature \TQ = 3.2 K and the N\'{e}el
temperature \TN = 2.4 K.\cite{Effantin1985} Phases I through III are 
labeled 
paramagnetic (PM), anti ferro quadrupolar (AFQ), and 
antiferromagnetic (AFM), respectively.}
\label{phasdiag}
\end{figure}

The dense Kondo system \cebsix~ (\TK $\sim$ 1 K) exhibits a three part
phase diagram (see Figure \ref{phasdiag}).  This paper reports new
high field measurements of the phase I to phase II transition 
temperature in the H-T plane,
\TQH.
Cerium hexaboride is one of several rare earth hexaborides that
crystallize in the primitive cubic structure with the rare earth
ions at the cube center and boron octahedra at the cube corners.  In
the past decade there have been many studies of the electronic,
thermal and magnetic properties of \cebsix~ because of interest in the
low temperature heavy fermion (HF) ground state.\cite{Schlottmann1989}
All of the magnetic properties arise from the single $4f$ electron on
the Ce atom that hybridizes with the conduction electrons to give rise
to the HF behavior.

The largest factor influencing the energy levels of the $4f$
electron on the Ce atom in \cebsix~ is the spin-orbit interaction.  This
interaction splits the 14-fold degenerate $4f$ level into a 6-fold
degenerate, $^{2}F_{5/2}$, and an 8-fold degenerate, $^{2}F_{7/2}$, level.  The $^{2}F_{5/2}$
level lies lowest in energy and is separated from the $^{2}F_{7/2}$ level by
an energy much greater than 500 K. Thus only the J = 5/2 state is
populated at room temperature and below.  In the absence of any other
effects the magnetic sublevels would correspond to J = $\pm$ 1/2, $\pm$
3/2 and $\pm$ 5/2 with a Lande g-factor for this level of 6/7.
 
Point ion crystal field theory \cite{Low1960} predicts that the cubic
crystal field due to the six borons in \cebsix~ further splits the Ce
6-fold degenerate $^{2}F_{5/2}$ level into a 2-fold degenerate
$\Gamma_{7}$ and a 4-fold $\Gamma_{8}$ level.  There have been
different interpretations of data with differing conclusions about the
energy ordering of these two
levels,\cite{Lee1972,Aoki1980,Hanzawa1981,Hanzawa1982} but it is now
generally perceived that in \cebsix~ the $\Gamma_{8}$ is the lowest
energy state, and the splitting between the $\Gamma_{7}$ and the
$\Gamma_{8}$ levels is on the order of 530 K.\cite{Zirngiebl1984} The
$\Gamma_{8}$ symmetry of the $f$ electron on Ce allows not only a
magnetic dipole moment, but in addition, an orbital electric and
magnetic quadrupole moment.  In zero applied magnetic field several
different orderings of these moments have been proposed to occur.

The overall results of the previously published magnetic field -
temperature phase diagram of \cebsix~ is shown in Figure 
\ref{phasdiag}.\cite{Effantin1985}  At high
temperatures the material is paramagnetic (Phase I) with 2.34 $\mu_{B}$ per
Ce atom.  In zero applied field, as the temperature is decreased,
there is a transformation into the first ordered state at 3.5 K (Phase
II), then at 2.2 K the Ce dipole moments align antiferromagnetically
(Phase III).  There are several substructures within Phase III, but we
will not be concerned with the structure of Phase III other than to
point out that at all applied magnetic fields above about 2.2 T it
does not exist.

The ordering in Phase II was studied by neutron diffraction and
proposed to be an ordering of quadrupole moments.\cite{Effantin1985}
Antiferro quadrupolar ordering has been observed in other materials, 
for example, TmTe.\cite{Matsumura1996}  In TmTe this AFQ ordering is 
destroyed by applied magnetic fields of higher than 6 T.
As can be seen from the published phase diagram for \cebsix, the state is not
destroyed by the application of magnetic fields up to 15 T. In this 
AFQ
model it is the coupling between the orbital quadrupole and spin
dipole moments that allows the phase transition to be observed with
magnetic torque measurements in uniform fields.

\section{Measurements}\label{sec:meas}
The magnetic measurements were carried out with a metal film
cantilever magnetometer, composed of two metal plates (one fixed and the
other flexible) that senses forces and torques
capacitively.  A single crystal of \cebsix~ is attached to the
flexible plate with Apiezon N grease.  When the sample/cantilever is 
positioned at field center, the sample experiences a torque 
proportional to its magnetization.  Most of the measurements reported 
here were made at field center.  However, three data points were taken 
with the sample in a field gradient (0.2 T/cm), where the sample 
experiences a force proportional to its magnetization.  The data is 
summarized in Figure \ref{tqphase}.

\begin{figure}[p]
	$$\epsffile{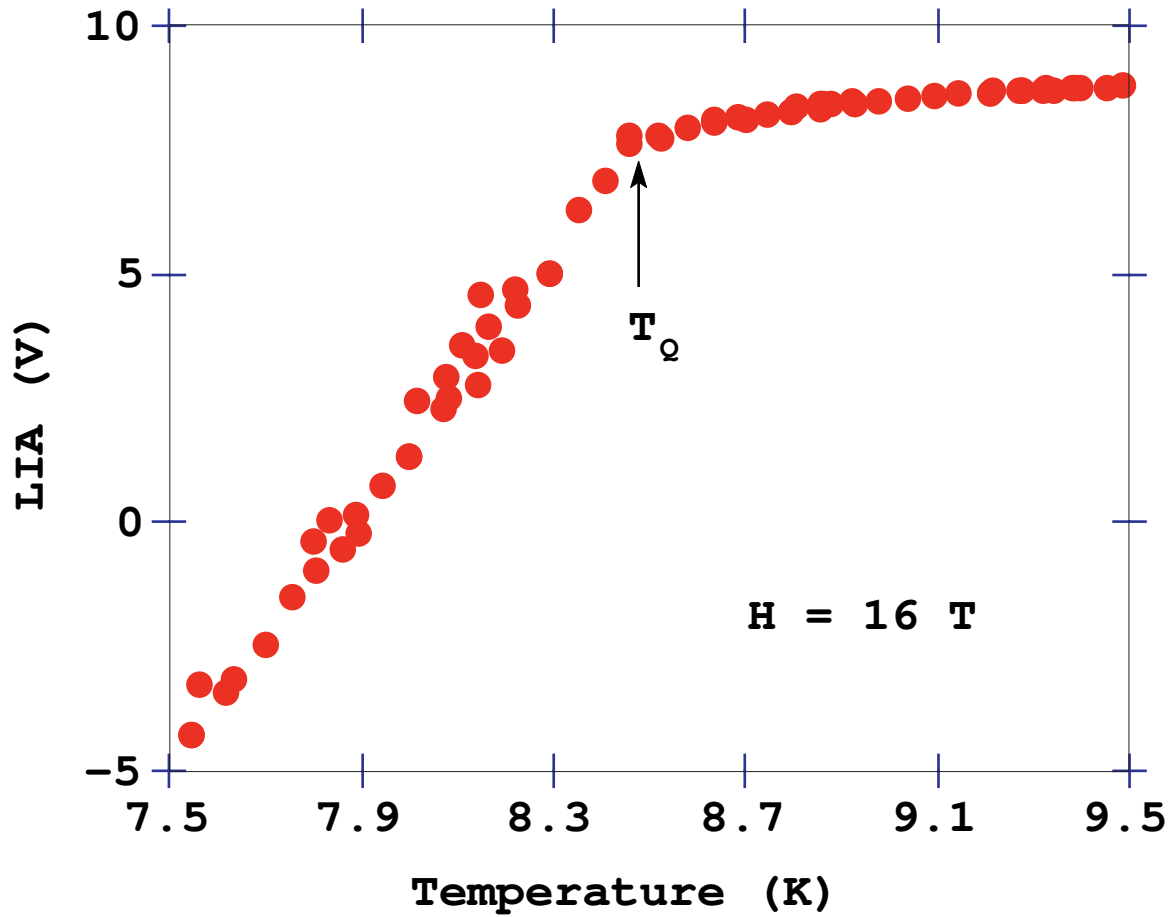}$$
\caption{This is an example of raw data taken at an applied magnetic 
field of 16 T.  The temperature was ramped slowly through the 
quadrupolar transition shown as \TQ~ on the plot.  The change in slope 
of the capacitance versus temperature curve is taken as \TQ.}
\label{rawdata}
\end{figure}

The sample's magnetization was measured at fixed fields as a function 
of temperature (as shown in Figure \ref{rawdata}).  To ensure proper 
determination of temperature a Lake Shore Cernox\tm~ CX 1030 series resistive 
thermometer was thermally anchored to the flexible plate of the 
cantilever with Cry-Con grease,\cite{Lakeshore} and corrections were 
made for the magnetic field 
dependence of the Cernox\tm~ thermometer.  Details of how such 
corrections should be made can be found in a paper by Brandt \etal
\cite{Brandt1999}

\section{Discussion}\label{sec:discuss}
Because of the antiferromagnetic ordering with wave vector k$_{0}$ =
[1/2, 1/2, 1/2] observed in neutron diffraction\cite{Effantin1985} the
ordering in Phase II was proposed to be that of quadrupole moments,
requiring a splitting of the four-fold degenerate $\Gamma_{8}$ ground
state into two doublets.  Several models have been given for this
splitting.  Either a dynamic Jahn-Teller effect involving acoustic
phonons, or a hybridization-mediated anisotropic coupling of the $4f$
wave functions to the $p$-like boron or $5d$-type cerium wave
functions were suggested as possibilities in Ref. 
\onlinecite{Zirngiebl1984}.  An alternative interpretation of these 
neutron scattering results has been given by Uimin in Ref. 
\onlinecite{Uimin1996c}.  Uimin interprets the low temperature 
frequency shift of the $\Gamma_{7}$ - $\Gamma_{8}$ as arising from 
collective modes of spin fluctuations caused by the orbital degrees of freedom.

In an early paper Ohkawa \cite{Ohkawa1983} proposed that indirect
exchange interactions between pairs of Ce atoms would produce a
splitting of the four-fold degenerate level into (4 $\times$ 4)
sixteen levels split into a group of two triplets and a group
consisting of a singlet plus a nine-fold degenerate level with Phase
II representing an ordering of the orbital moments.  Calculations in
Ref.  \onlinecite{Ohkawa1983} based on this model predict that the
critical field that destroys Phase II will be in excess of 30 - 50 T.

Building on Ohkawa's work, Shiina \etal \cite{Shiina1997} have 
constructed a mean field theory for Ohkawa's RKKY model and calculated 
the phase diagram.  They argue that the increase of \TQH~ at low fields 
is due mainly to field-induced dipolar and octupolar moments.  Also, 
they suggest an improvement to the model by 
introducing asymmetry into the interaction between dipolar and 
octupolar moments which leads to induced staggered dipolar moments 
and accounts for the distinction of Phase II into a low field phase 
and a high field phase suggested by Nakamura \etal 
\cite{Nakamura1995}.  However, detailed measurements on the symmetry of 
the order parameter are required to see what applicability Shiina 
\etal 's model has to \cebsix.

Uimin \cite{Uimin1996b,Uimin1997} described the shape \TQH~ as 
arising from competing AFQ patterns near the ordering temperature.  
These fluctuations are suppressed by an applied magnetic field.  
Uimin's model predicts three important characteristics of the 
AFQ-Paramagnetic
phase diagram: (1) that \TQH~ increases linearly at low applied 
fields, (2) that the AFQ-Paramagnetic phase line is anisotropic in the 
H-T
plane, and (3) that \TQH~ decreases and goes to zero at sufficiently 
high fields.  Based on data available at the time \cite{Milton} Uimin
estimated the lower limit field for
the re-entrance of \TQH~ as approximately 25 - 30 T yielding an H(\TQ~ = 0)
approaching 80 T.\cite{Uimin1997}  The
measurements reported here do not show re-entrance up to 30 T.  Uimin 
points out that his estimate of H(\TQ~ = 0) does not take into account 
the Kondo effect; however, the measurements are carried out at higher 
energies than the Kondo energy (on the order of 2 K).

Uimin's theoretical treatment also found a significant dependence of 
\TQH~ on the orientation 
of the applied field.  In the [111] \TQH~ does not decrease for 
arbitrarily high fields.  Our measurements were made with the sample 
in the [100].  However, no experiment has shown any significant 
orientation dependence in the \TQH~ phase line, which Uimin attributes 
to the unusual anisotropy of the Zeeman energy.

More recently, Kasuya \cite{Kasuya1998} has considered a paired dynamic Jahn-Teller
distortion with no quadrupolar ordering causing an increased Ce - Ce
antiferromagnetic (AFM) coupling that is enhanced by increasing
applied magnetic field.  In Ref. \onlinecite{Kasuya1998}  the
critical field at which this enhanced AFM ordering is destroyed also
is predicted to be greater than 30 T.

It should be noted that muon spin rotation measurements in zero 
applied magnetic field yield a different magnetic structure for 
\cebsix~ for both Phase II and III.\cite{Feyerherm1994,Feyerherm1995}  
Detailed measurement of the variation of magnetic order parameter as a 
function of temperature are needed.

\section{Conclusions}\label{sec:conclu}
\begin{figure}[p]
	$$\epsffile{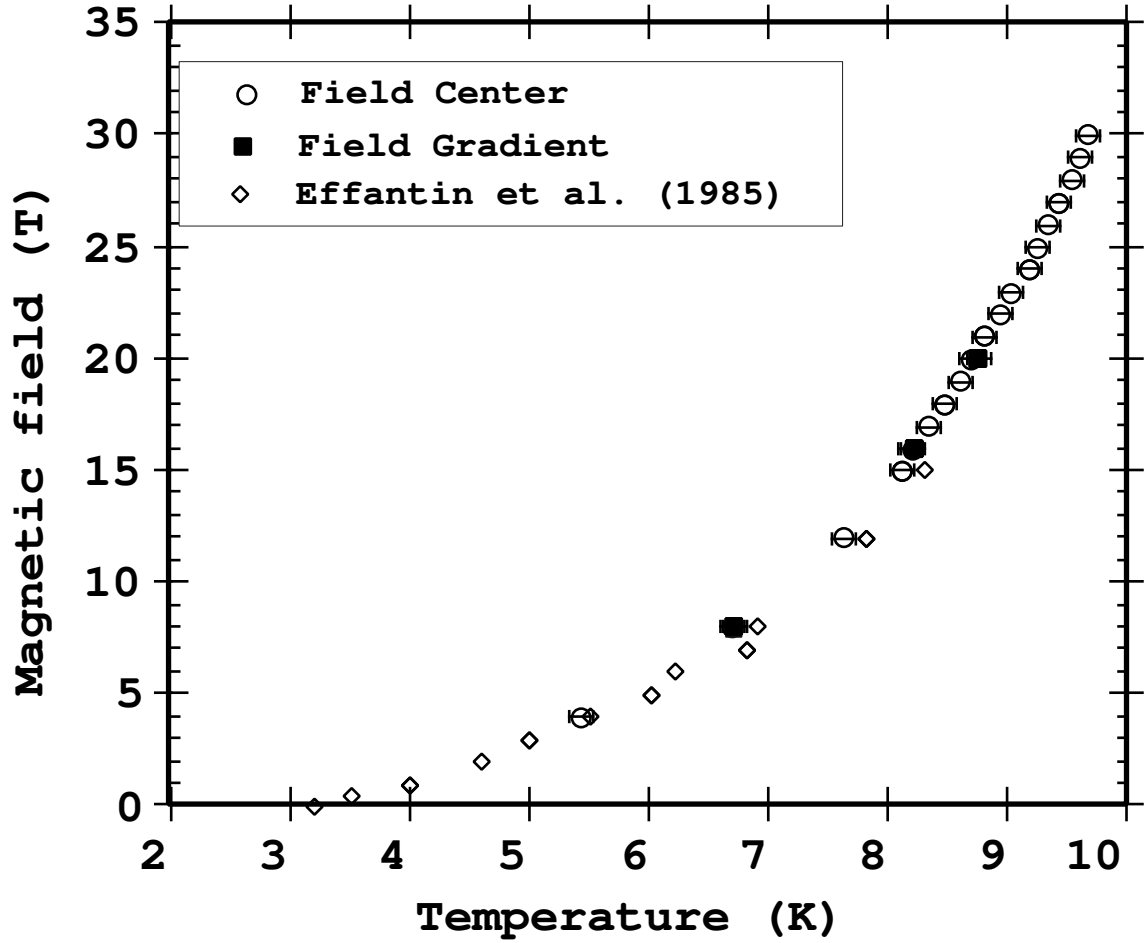}$$
\caption{The quadrupolar transition temperature \TQH~ is shown as a 
function of magnetic field.  New data are compared with 
previously published data.}
\label{tqphase}
\end{figure}

Our measurements of the phase boundary between Phase I and Phase II,
along with previously published points are shown in Fig. \ref{tqphase}.  As can
be seen, the present measurements below 15 T are in good agreement
with published values and double the measured field range.  The
slope of the phase boundary continues to increase with applied field
and becomes nearly independent of temperature above 25 T. There is no
indication that the phase is being destroyed with field up to 30 T. In
addition to measurements in uniform fields, we have included several
points that were taken in the presence of a strong magnetic field
gradient dH/dz, where both H and z are along the [100] axis of the
sample.  If Phase II includes antiferro ordering of magnetic
quadrupole moments, then the application of a field gradient should
exert a force on the moments causing them to align and destroy the
phase.  As can be seen the magnetic field gradient has no effect on
the transition temperature (see Fig. \ref{tqphase}).

In conclusion, it is seen that any theory that predicts the
destruction of Phase II below 30 T does not include either all of the
effects, or includes incorrect mechanisms.  Two theories presented to
date\cite{Ohkawa1983,Kasuya1998}, both of which are predicated on
indirect exchange, predict destruction of the phase at fields $>$ 30
T, and cannot be ruled out.  Additional measurements would aid in
distinguishing between competing theories of the magnetically ordered
phase.  Clearly, the phase diagram \TQH~ needs to be measured to high
fields.  Also, measurements on the alloy series \celabsix~ will assist
in understanding the splitting of the $\Gamma_{8}$ level as the Ce
concentration increases.

This work was supported in part by the National Science Foundation
under Grant No.  DMR-9971348 (Z. F.).  A portion of this work was
performed at the National High Magnetic Field Laboratory, which is
supported by NSF Cooperative Agreement No.  DMR-9527035 and by the
State of Florida.


\onecolumn
\widetext

\newpage 

\newpage 

\newpage 

\end{document}